\documentclass[a4paper]{spie}  

\usepackage{amsmath,amsfonts,amssymb}
\usepackage{graphicx}
\usepackage{upgreek}
\usepackage[colorlinks=true, allcolors=blue]{hyperref}

\newcommand*\aap{A\&A}

\newcommand*\ao{Appl Opt}

\newcommand*\apjl{ApJ}

\newcommand*\apjs{ApJS}

\usepackage{amsmath,amsfonts,amssymb}
\usepackage{graphicx}
\usepackage[colorlinks=true, allcolors=blue]{hyperref}
\usepackage{threeparttable} 
\usepackage{macro_ref}
\usepackage{geometry}
\title{L-band nulling interferometry at the VLTI with Asgard/Hi-5: status and plans} 

\author[a]{Denis Defrère}
\author[a]{Azzurra Bigioli}
\author[b]{Colin Dandumont}
\author[a]{Germain Garreau}
\author[a]{Romain Laugier}
\author[a]{Marc-Antoine Martinod}
\author[b]{Olivier Absil}
\author[c]{Jean-Philippe Berger}
\author[d]{Emilie Bouzerand}
\author[e]{Benjamin Courtney-Barrer}
\author[f]{Alexandre Emsenhuber}
\author[g,h]{Steve Ertel}
\author[i]{Jonathan Gagne}
\author[d]{Adrian M. Glauser}
\author[h]{Simon Gross}
\author[k]{Michael J. Ireland}
\author[k]{Harry-Dean Kenchington}
\author[a]{Jacques Kluska}
\author[l]{Stefan Kraus}
\author[m]{Lucas Labadie}
\author[b]{Viktor Laborde}
\author[n]{Alain L\'eger}
\author[h]{Jarron Leisenring}
\author[b,o]{Jérôme Loicq}
\author[c]{Guillermo Martin}
\author[a]{Johan Morren}
\author[p]{Alexis Matter}
\author[b]{Alexandra Mazzoli}
\author[a]{Kwinten Missiaen}
\author[a]{Muhammad Salman}
\author[n]{Marc Ollivier}
\author[a]{Gert Raskin}
\author[b,h]{Hélène Rousseau}
\author[m,j]{Ahmed Sanny}
\author[a]{Simon Verlinden}
\author[a]{Bart Vandenbussche}
\author[d]{Julien Woillez}

\affil[$a$]{Institute of Astronomy, KU Leuven, Celestijnenlaan 200D, 3001, Leuven, Belgium}
\affil[$b$]{Space sciences, Technologies \& Astrophysics Research (STAR) Institute, University of Li\`ege, Li\`ege, Belgium}
\affil[$c$]{Univ. Grenoble Alpes, CNRS, IPAG, 38000 Grenoble, France}
\affil[$d$]{Eidgen\"ossische Technische Hochschule (ETH) Zurich, Institute for Particle Physics and Astrophysics, Zurich, Switzerland}
\affil[$e$]{European Southern Observatory, Karl-Schwarzschild-Straße 2, 85748 Garching, Germany}
\affil[$f$]{Ludwig-Maximilians-Universität München, Germany}
\affil[$g$]{Large Binocular Telescope Observatory, 933 North Cherry Ave, Tucson, AZ 85721, USA}
\affil[$h$]{Department of Astronomy and Steward Observatory, 933 North Cherry Ave, Tucson, AZ 89 85721, USA}
\affil[$i$]{Université de Montréal, Canada}
\affil[$j$]{Macquarie University, Australia}
\affil[$k$]{Research School of Astronomy and Astrophysics, Australian National University, Canberra, ACT 2611, Australia}
\affil[$l$]{School of Physics and Astronomy, University of Exeter, Exeter, United Kingdom}
\affil[$m$]{I. Physikalisches Institut, Universit\"at zu K\"oln, Z\"ulpicher Str. 77, 50937 Cologne, Germany}
\affil[$n$]{Institut d'Astrophysique Spatiale, Universit\'e Paris Sud}
\affil[$o$]{Delft University of Technology, Netherlands}
\affil[$p$]{Laboratoire Lagrange, Universit\'e C\^ote d'Azur, Observatoire de la C\^ote d'Azur, CNRS, Boulevard de l'Observatoire, CS 34229, 06304, Nice, France}

\authorinfo{Further author information: Denis Defr\`ere: E-mail: denis.defrere@kuleuven.be}

\begin{document} 
\maketitle

\begin{abstract}
Hi-5 is the L’-band (3.5-4.0 $\mu$m) high-contrast imager of Asgard, an instrument suite in preparation for the visitor focus of the VLTI. The system is optimized for high-contrast and high-sensitivity imaging within the diffraction limit of a single UT/AT telescope. It is designed as a double-Bracewell nulling instrument producing spectrally-dispersed (R=20, 400, or 2000) complementary nulling outputs and simultaneous photometric outputs for self-calibration purposes. In this paper, we present an update of the project with a particular focus on the overall architecture, opto-mechanical design of the warm and cold optics, injection system, and development of the photonic beam combiner. The key science projects are to survey (i) nearby young planetary systems near the snow line, where most giant planets are expected to be formed, and (ii) nearby main sequence stars near the habitable zone where exozodiacal dust that may hinder the detection of Earth-like planets. We present an update of the expected instrumental performance based on full end-to-end simulations using the new GRAVITY+ specifications of the VLTI and the latest planet formation models. 
\end{abstract}


\section{Introduction}
\label{sec:intro}  

One of the main challenges to achieve high-contrast interferometric observations is to accurately remove the overwhelmingly dominant flux of the host star from the scientific signal\cite{defrere_review_2020}, similar to coronagraphy in single-pupil direct imaging. Over the last twenty years, a series of nulling interferometers\cite{bracewell_detecting_1978} have been deployed on state-of-the-art facilities, both across single telescopes and as separate aperture interferometers\cite{spalding_unveiling_2022}. These include the BracewelL Infrared Nulling Cryostat\cite{hinz_imaging_1998}, the Keck Interferometer Nuller\cite{serabyn_keck_2012}, the Palomar Fiber Nuller\cite{mennesson_high-contrast_2011,serabyn_nulling_2019}, the Large Binocular Telescope Interferometer\cite{hinz_overview_2016}, and DRAGONFLY/GLINT on Subaru/SCExAO\cite{norris_first_2020}. Considering the most recent three instruments (i.e., GLINT, PFN, and LBTI), the use of nulling interferometry allowed to gain one order of magnitude on the final post-processed contrast levels down to $\sim$10$^{-4}$. This can be explained theoretically by the fact that error terms linear in phase and/or amplitude are present at both peak and quadrature, but all linear error terms vanish at null, leaving only smaller quadratic error terms\cite{serabyn_nulling_2019}. The high null depth accuracies obtained with nulling interferometers were also made possible thanks to a combination of factors: the ability to use single-mode fibers (PFN) or integrated optics (GLINT), the use of the telescope's extreme adaptive optics system as a cross-aperture fringe tracker, and the introduction of a significantly improved technique for null-depth measurement, i.e., \emph{Numerical self calibration}\cite{hanot_improving_2011,mennesson_high-contrast_2011,defrere_nulling_2016,norris_first_2020,martinod_scalable_2021}. Much was learned about instrumental limitations with the scientific exploitation of these instruments. High-sensitivity mid-IR instruments such as the LBTI is mostly limited by the high thermal background radiation and the excess low frequency noise associated with the detector. At shorter wavelength, where the thermal background is less of an issue, the main limitations are related to high-frequency phase fluctuations and polarization errors. These limitations currently make state-of-the-art nulling interferometer operate one to two orders of magnitudes above the fundamental photon noise limit\cite{colavita_keck_2009,defrere_nulling_2016}.\\

An unexplored and very promising wavelength range for high-performance nulling interferometry is the L band (3 to 4\,$\mu$m). This wavelength range is also key for various science programmes on young exoplanetary systems and exozodiacal disks. In this paper, we present the status of Hi-5, a high-contrast L-band nulling interferometric instrument currently being developed for the visitor focus of the VLTI\cite{defrere_path_2018,defrere_interferometric_2018,defrere_hi-5:_2018} as part of the Asgard instrument suite\cite{asgard_martinod}. By leveraging its state-of-the-art infrastructure, ongoing GRAVITY+ upgrades, long baselines, and strategic position in the Southern hemisphere, a dedicated high-contrast VLTI instrument will be able to carry out several exoplanet programmes to study young Jupiter-like exoplanets at the most relevant angular separations\cite{fernandes_hints_2019,fulton_california_2021} (i.e., close to the snow line) and better understand how planets form and evolve. The Hi-5 project received fundings from OPTICON for a preliminary study and was later funded by the European Research Council (2020-2025). In this paper, we give an update of the project, its science case and current design.

\section{Hi-5 science case}
\label{sec:science}

\subsection{Exoplanet Formation Entropy}

\begin{figure*}
\begin{center}
{\includegraphics[width=0.95\textwidth]{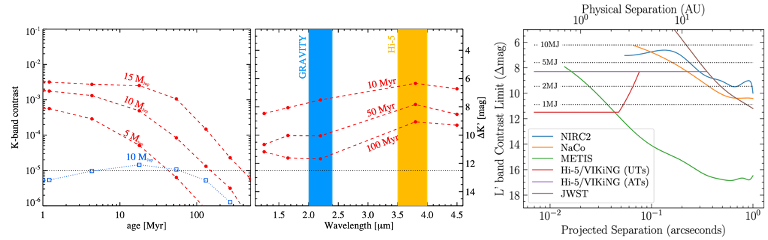}
      \caption{Left: Predicted K-band planet-star contrast for a Sun-like star as a function of age and given for different planet masses (red circles, BT-Settl models). The lower blue line shows the contrast predicted for the ``cold start" model. Middle: contrast between a 10-M$_{\rm J}$ planet and a Sun-like star for three different ages as a function of wavelength. Right: Contrast limits for a planet around a typical L’=7 apparent magnitude moving group star\cite{wallace_constraints_2021}. The total number of directly detected planets with accurate masses is expected to be 8$\pm$3 – roughly the same as METIS.}
\label{Fig:contrast}}
\end{center}
\end{figure*}

\begin{figure*}
\begin{center}
{\includegraphics[width=0.95\textwidth]{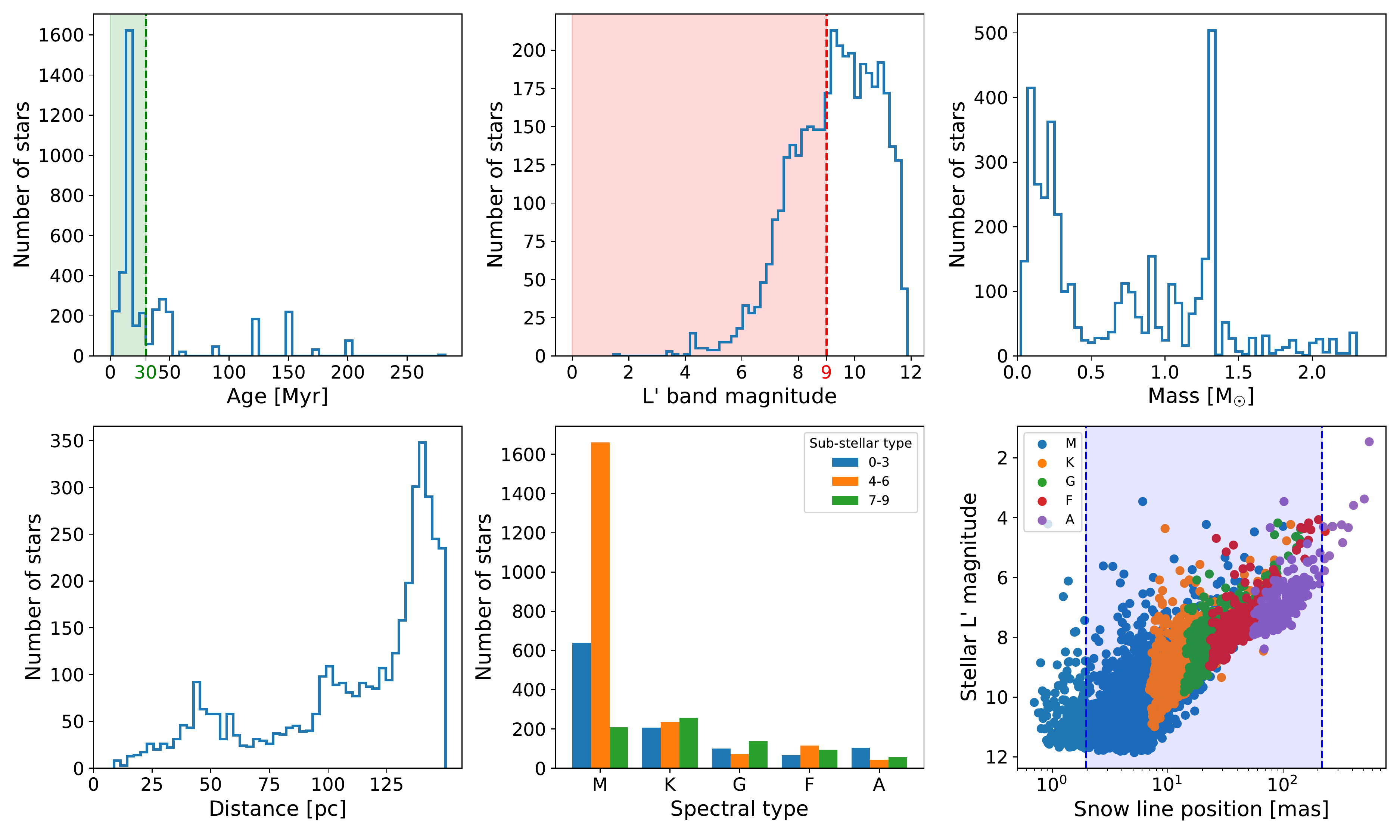}
      \caption{Age, L'-band magnitude, mass, distance, spectral-type, and snowline position distribution for our stellar catalog\cite{hi5_dandumontA}. The catalog is composed of $4002$ stars. All of these stars are thought to be in young moving groups. The green vertical line in the age histogram corresponds to the planet/stellar age when the cooling of the planet prevents to distinguished cold from hot-start formation model\cite{wallace_constraints_2021}. The red vertical line in the magnitude histogram corresponds to the performance of \textsc{NAOMI}\cite{woillez_naomi_2019}. The blue zone in the snowline distribution represents the FoV of the ATs (IWA-OWA).}
\label{Fig:StellarCatalogue}}
\end{center}
\end{figure*}

The scarcity of giant exoplanets discovered by single-aperture direct imaging surveys currently challenges our understanding and theories of planet formation. The GAIA spacecraft is expected to measure the masses of many thousands of giant exoplanets \cite{perryman_astrometric_2014}, of which tens are expected to be in nearby young moving groups \cite{wallace_likelihood_2019}. The age of groups and stars within them are also in the process of being more tightly constrained by GAIA data \cite{Crundall2019}. When these new data sets are combined with a direct detection of the planets using Hi-5, the formation mode of the planets can be directly constrained \cite{wallace_likelihood_2019}. Low-resolution spectroscopic observations of such planets in the L band will provide their radius and effective temperature as well as critical information to study the non-equilibrium chemistry of their atmosphere \cite{skemer_first_2016}. By tracing the thermal emission from these exoplanets, opacities of absorbers (molecular bands, dust, clouds), vertical temperature profiles, atmospheric dynamics, rotation rate, and formation processes of giant exoplanets can be constrained \cite{skemer_first_2016}. This is often characterized by the formation entropy as a function of planetary mass, which splits into hot-start and cold-start models (enabled by gravitational instability and core accretion respectively). The largest differences are for the more massive planets, which are also easier to detect (see Figure~\ref{Fig:contrast}). Finding the dominant mode of giant planet formation as a function of key parameters such as planetary mass and distance from the host star is one of the most critical unknown questions in planetary formation. As shown in Figure~\ref{Fig:contrast}, Hi-5 is not only anticipated to be more powerful in detecting these key exoplanets with mass measurements than existing instruments, it is also expected to be more powerful than METIS in a key region of parameter space (giant planets of $\sim$1\,au separation, corresponding to 20 mas at 50\,pc. The total observing time will be limited by the number of GAIA mass measurements, which is expected to be of order 24 \cite{wallace_likelihood_2019}. An updated target catalog based on the latest GAIA data release is shown in Figure~\ref{Fig:StellarCatalogue}\cite{hi5_dandumontA}.

\subsection{Exoplanet in transition disks}

Understanding the formation of exoplanets requires observing their formation within protoplanetary disks around young stars. Spatially resolved observations with ALMA and direct imaging instruments like SPHERE have revolutionized our view of protoplanetary disks. While they were modeled as axisymmetric and continuous, they turned out to be very structured with the presence of rings, gaps, spirals and large cavities. Disks with large cavities are called transition disks. The observed cavity is often large (>20 au) and is observed in 10\% of the disks \cite{vanderMarel2018}. Many of these cavities also show spectacular asymmetric substructures such as spirals or vortices \cite{vanderMarel2021}. The origin of these cavities is generally linked to the presence of companions that carve the cavity in both dust and gas \cite{Norfolk2021}. The planet-disk interactions would also explain the azimuthal perturbations witnessed in these disks \cite{Boccaletti2020}. Nevertheless, only two accreting planets were detected so far in one such transition disk system (e.g., PDS 70 \cite{Keppler2018,Haffert2019}). For other transition disks, the detection limits from direct imaging are not sufficient to confirm the planetary origin of the cavity \cite{vanderMarel2021b} because of the lack of sensitivity inside the cavity, close to the star. Importance to the field: We propose to take advantage of the high angular resolution and sensitivity of Hi-5 to look for planets in the transition disk cavity, to be able to estimate their masses and their positions. Based on the expected sensitivity and contrast performances of Hi-5, we estimate the detection limits to reach the planetary regime down to 1 M$_{\rm J}$ for the youngest systems ($\sim$2 Myr). These observations will provide an unprecedented input for hydrodynamical models to understand planet-disk interaction and constrain planet formation and evolution in the very early stages. The high-contrast nulling mode of Hi-5 combined with the long VLTI baselines will provide the dynamic range and angular resolution required to observe the cavities of transition disks imaged at shorter wavelengths by current direct imaging instrument such as SPHERE and currently inaccessible by other L-band single-dish exoplanet imagers. 	
\subsection{The prevalence and nature of exozodiacal disks}

Exozodiacal dust (exozodi) is warm and hot dust in the inner regions of main-sequence planetary systems, including the Habitable Zone (HZ). It presents both a scientific opportunity and a programmatic obstacle for future exo-Earth imaging. Programmatically, light from HZ dust adds photon noise and disk structures cause confusion \cite{defrere_direct_2012}. This puts make-or-break constraints on our ability to detect and characterize habitable exoplanets. Since this is a major goal of the worldwide astronomical community – possibly requiring the bundling of the majority of the astronomy funding for decades to come – exozodi science critically contributes to guiding the path of all astronomy for a whole century. This was prominently highlighted in ESA’s Voyage 2050 and NASA’s 2020 Decadal Survey. Scientifically, the dust is the most readily detectable constituent of a star's HZ environment through direct methods. Observations of the dust can be used to trace the presence and dynamics of its parent bodies (comets, asteroids) and planets. This allows us to study the environment in which rocky, habitable-zone planets exist. Cometary and asteroid impacts over its lifetime critically add to – or remove – a planet’s atmosphere, depending on the impact energy and impactor composition \cite{Wyatt2020}, thus affecting their potential for actually being habitable as compared to, e.g., barren or water worlds. Studying exozodiacal dust thus helps locate habitable planets in the Universe and understand their prevalence.

\begin{figure*}
\begin{center}
{\includegraphics[width=0.95\textwidth]{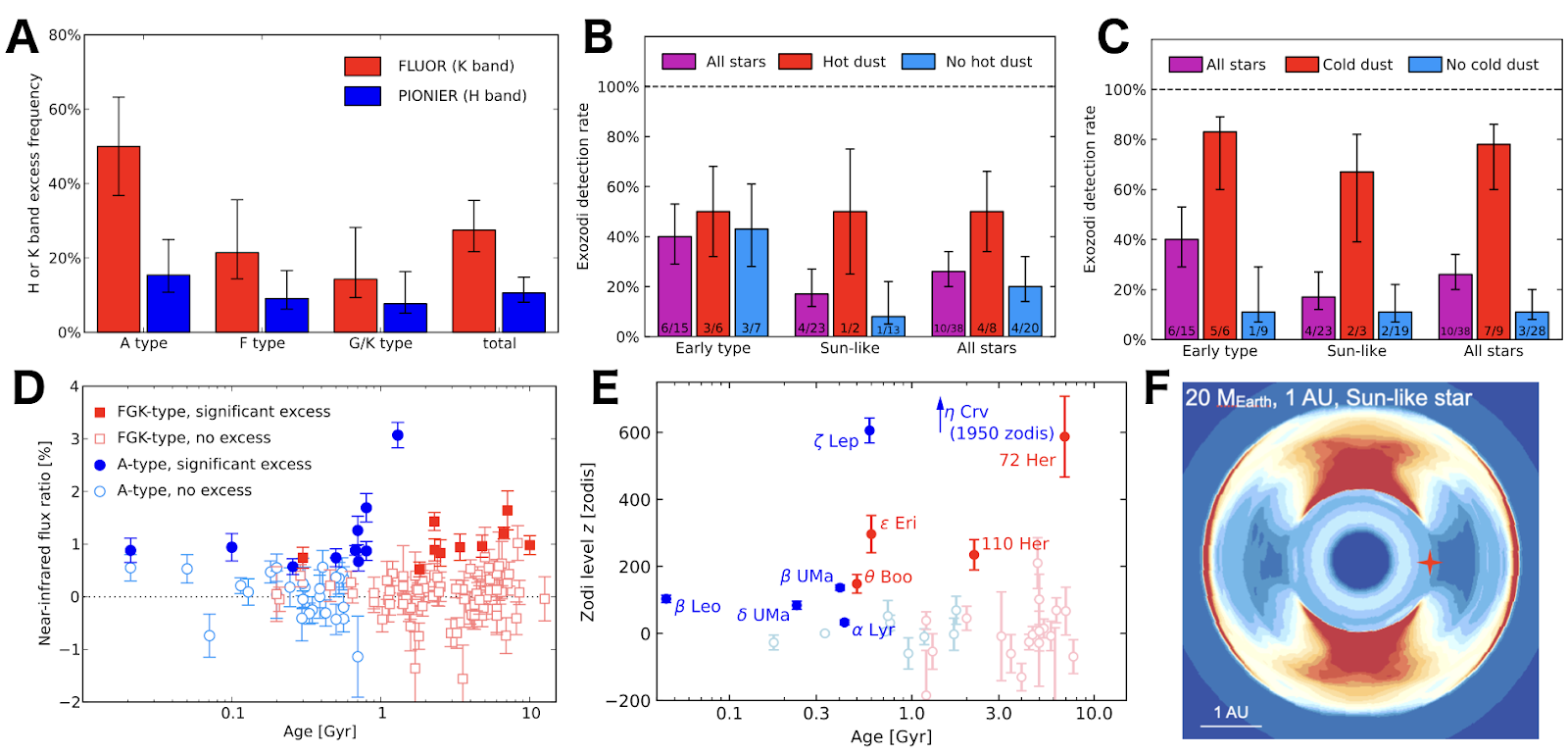}
      \caption{Critical progress in understanding exozodiacal dust has been made from our previous interferometric surveys, however, conclusions are critically limited by the small number of detections and poor S/N due to limited sensitivity that are overcome with Hi-5. (A) Detection statistics of hot dust vs. wavelength and stellar spectral type.  (B) Detection statistics of HZ dust vs. stellar spectral type and the presence of hot dust.  (C) Detection statistics of HZ dust vs. spectral type and presence of a cold debris disk.  (D) NIR and (E) N-band excess measurements (detections filled, none-detections empty symbols) vs. stellar age and spectral type.  (F) Simulated structures in a HZ dust disk from interaction with a super-Earth/mini-Neptune mass planet orbiting at 1 au. References: \cite{ertel_near-infrared_2014,ertel_hosts_2020,Pearce2020}}
\label{Fig:exozodi}}
\end{center}
\end{figure*}

Exozodiacal dust is so far only observed in thermal emission due to the lower dust-to-star contrast compared to scattered light, but the dust’s excess emission is still too faint to detect photometrically. It thus needs to be spatially resolved from the star, and the small angular scales and high contrasts involved (1 au at 10 pc corresponds to 0.1”) require the use of precision interferometry. We have used VLTI/PIONIER and CHARA/FLUOR to detect hot exozodiacal dust in the near infrared (nIR, \cite{absil_near-infrared_2013,Absil2021,ertel_near-infrared_2014,ertel_near-infrared_2016}) and the Large Binocular Telescope Interferometer (LBTI) to detect truly HZ dust in the N-band \cite{ertel_hosts_2018,ertel_hosts_2020}. Our surveys have provided critical statistical information on the dust’s origin and correlation with the mass and age of the host star and the presence of a cold debris disk (see Figure~\ref{Fig:exozodi}). These observations have given rise to our own \cite{Bonsor2012,Bonsor2013,Bonsor2014,Faramaz2017,Rigley2020} and community-driven \cite{vanLieshout2014,Pearce2020} theoretical studies of exozodis. However, due to the limited signal-to-noise ratio (S/N) and single-wavelength nature of the available detections, reaching a general understanding of exozodis has so far been elusive.

Nulling interferometry on the VLTI will provide a tenfold better sensitivity to extended circumstellar emission than the more common, constructive optical long-baseline interferometry used with PIONIER and FLUOR. Furthermore, our observations using the limited sensitivity of MATISSE have shown that PIONIER detections in H-band have a fivefold more favorable dust-to-star flux ratio in L-band for a fifty-fold total S/N improvement. This S/N improvement for exozodi observations over previous instruments will allow us to finally overcome constraints from low-significance detections and small-number statistics to the point where a true exozodi luminosity function can be derived and correlated with host star and planetary system properties. Hi-5’s observing wavelength critically connects the nIR and N-band detections, thus allowing for creating a unified picture of exozodiacal dust across all relevant spatial scales and observing wavelengths. Ultimately, the high S/N of the Hi-5 observations will likely allow for image reconstruction and thus the detailed study of the dust origin through its distribution and the search for structures indicative of planet-disk interaction.

We plan to survey a volume- and brightness-limited sample of $\sim$100 nearby stars (30 nights) with the ATs analog to our previous, extremely successful survey with VLTI/PIONIER. Our sample is largely unbiased in several core properties such as stellar spectral type and age, and the presence of a known, cold debris disk. In addition, the sample includes stars with previous detections of exozodiacal dust at both near- and mid-infrared wavelengths. Follow-up observations for the detailed characterizing the highest S/N detections will be proposed separately from the core survey. An analysis of all Hi-5 data searching for or characterizing exoplanets around mature stars will also be performed to search for exozodiacal dust and our exozodi observations can to some extent be used to search for companions.  While the observing strategies for exoplanets and exozodis are not exactly the same, significant information can be extracted from either data for both science cases.

\section{Instrumental overview}
\subsection{Asgard and instrumental components}

\begin{figure}[b]
    \centering
    \includegraphics[width=0.95\textwidth]{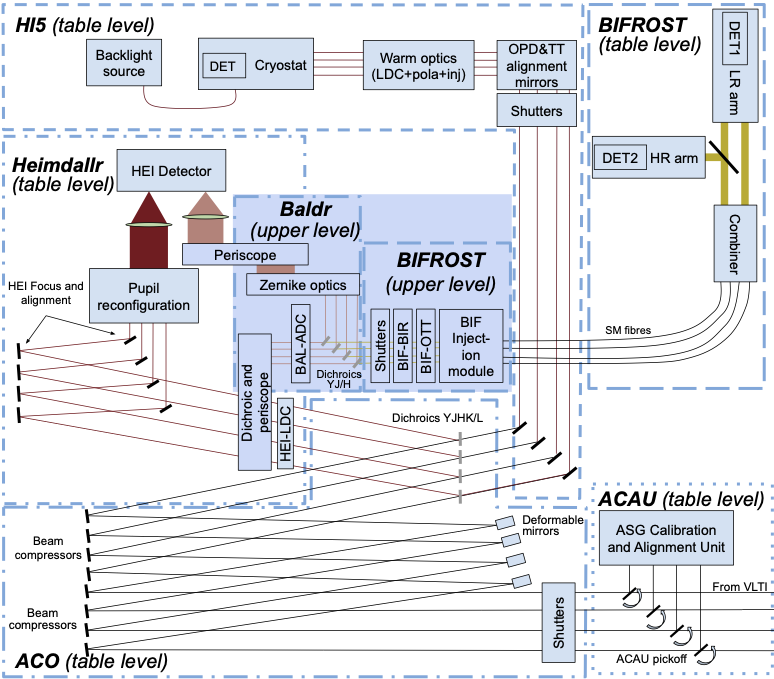}
    \caption{Preliminary conceptual layout of Asgard and its different modules (not to scale).}
    \label{fig:asgard_diagram}
\end{figure}

The current system block diagram of Asgard is shown in Figure \ref{fig:asgard_diagram}. The four beams from the VLTI telescopes enter the diagram at the bottom right, go through shutters, and then the deformable mirrors. The beam compressors reduce the pupil size from 18\,mm to 12\,mm. A dichroic then sends the L-band to Hi-5. The Hi-5 sub-systems are listed below. More information on the integration of these subsystems in Asgard and on Paranal can be found in a dedicated paper in these proceedings \cite{asgard_martinod}.

\begin{itemize}
    \item Beam conditioner (HI5-SHUT): comprises shutters and diaphragms. It shuts the beams individually for calibration purposes and equalises the intensities of the beams to optimise the null depth.
    \item OPD \& TT alignment mirrors (HI5-OTT): to cophase the input beams between Hi-5 and Asgard and to optimise the light injection into the photonic chip. They are actuated by stepper motors from Thorlabs.
    \item Warm optics (HI5-WO): comprises LDCs, polarisation control optics, the slicer and alignment camera. The LDC compensates for the chromatic phase between the K and L bands and the water vapour effects via a feedback loop. The polarisation control compensates for polarisation effects which reduce the instrumental null depth. The slicer overlaps the pupils on the cold stop in the cryostat. The alignment camera is an Infratec ImageIR 5300.
    \item Cryostat (HI5-CRYO): comprises a window, a cold stop, the photonic chip which combines the beams and creates photometric outputs, a wheel and imaging optics, the camera, the cryocooler and the vacuum system. The wheel contains grisms with different spectral resolutions. The chip is mounted on a 3-axe mount for alignment and focusing with HI5-WO. The cryocooler is the PT805 from Absolut system and the vacuum pump is the HiCube 30 Eco from Pfeiffer vacuum.
    \item Camera (HI5-DET): Camera (HI5-DET): comprises the detector and its electronics. It collects the spectrally-dispersed signal delivered by the photonic chip. It consists of Teledyne’s 5-micron Hawaii-2RG with SIDECAR cold electronics and Astroblank MACIE warm electronics.
    \item Backlight source (HI5-BACK): Point-like source used to align Hi-5 with Asgard.
\end{itemize}

\subsection{Current design of warm optics}

\begin{figure}[!h]
    \centering
    \includegraphics[height=0.95\textheight]{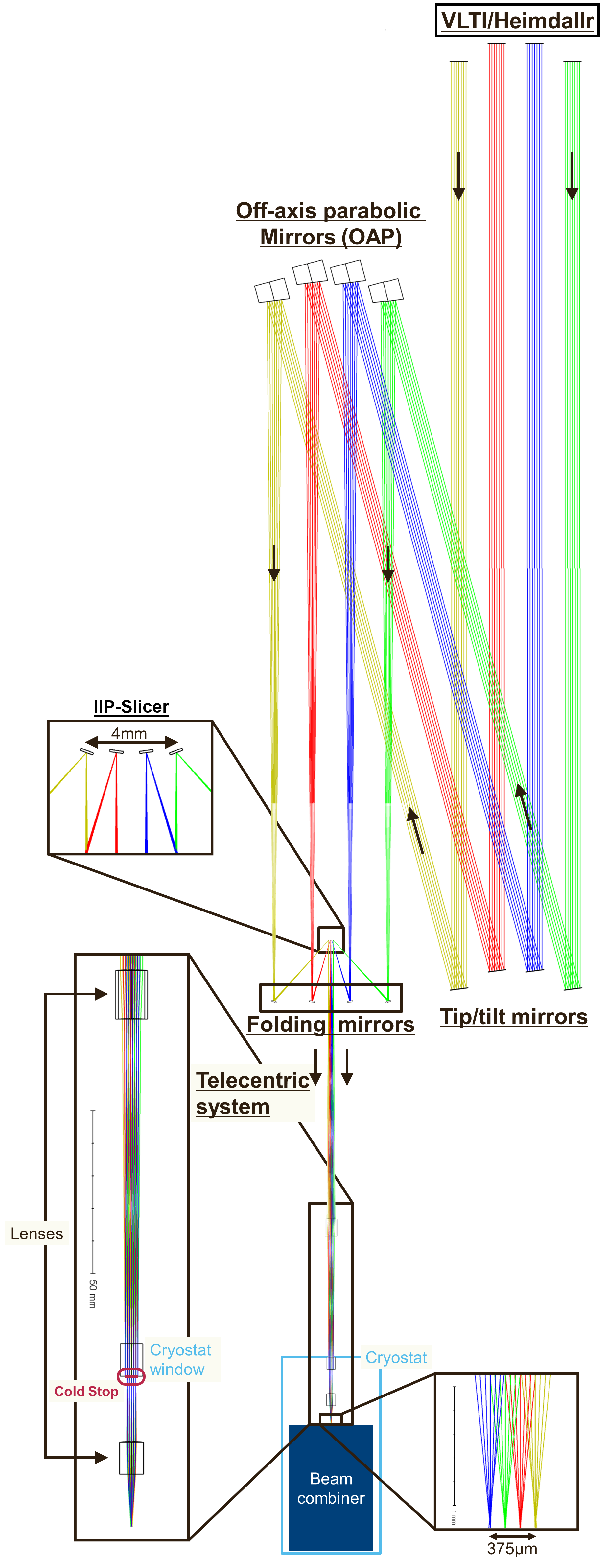}
    \caption{Design of the injection system on Zemax considered in this study.}
    \label{fig:injection system}
\end{figure}

In this section, we provide a brief description of the current optical design of the instrument. The optical design of the warm optics is shown in Figure\,\ref{fig:injection system}. The four beams coming from the VLTI arrive parallel and collimated with a diameter of 12\,mm after the HEIMDALLR fringe tracker and are focused via four identical Off-Axis Parabolas (OAP) onto an Intermediate Image Plane (IIP). All the design is implemented to be highly symmetric, in order to not add polarization and phase mismatch between the four arms. The focal length of these OAP will be customized to respect our surface quality standard. More information, including a tolerancing analysis, can be found in a dedicated paper in these proceedings\cite{hi5_garreau}. 

\newpage

\subsection{Current design of cold optics}

\begin{figure}[!h]
\begin{center}
\begin{tabular}{c}
\includegraphics[width=0.95\textwidth]{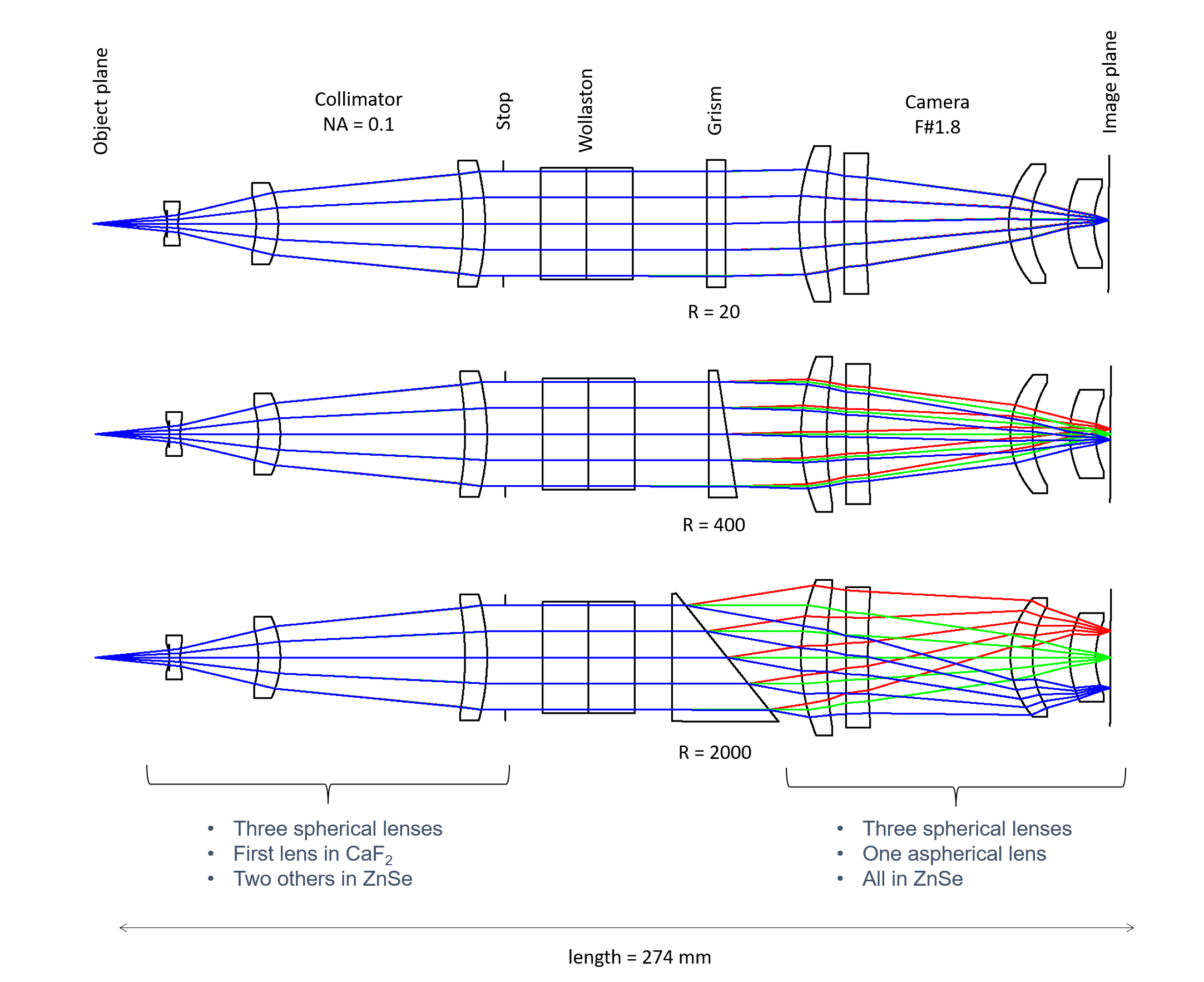}
\end{tabular}
\end{center}
\caption{\label{Fig:Fullspectro} Side view (cut) of the spectrometer optical layout. Each spectral resolution ($R = 20$, $400$, and $2000$) is depicted. Only the three germanium grisms are changed between each configuration (the focal plane is fixed). They will be mounted on a wheel mechanism.}
\end{figure}

Figure~\ref{Fig:Fullspectro} shows the optical design of the cold optics for the three different spectral resolutions ($R = 20$, $400$, and $2000$). The 8 outputs of the beam combiner are shown on the left. The numerical aperture of the collimator is related to the beam combiner. This collimator is made of three spherical lenses, the first one being in CaF$_2$ and the two others in ZnSe. Both materials are transparent in the L'-band. To split the polarization, a Wollaston is positioned just after. It could be removed without changing the focal plane (collimated beam + parallel faces). Three germanium grisms were developed and optimized. They will be mounted on a wheel mechanism. A change of spectral resolution will not affect the image plane position. Since the spectrometer is inside the cryostat, a change of spectral resolution will not need any technician intervention. The imaging part is composed of three spherical lenses and one aspherical lens. They are all in ZnSe. The detector, the HAWAI 2RG\textregistered, is located at the image plane. More information on the requirements, tolerancing, trade-off and optimization can be found in a dedicated paper in these proceedings\cite{hi5_dandumontB}. A separate paper also describes the current status of the photonic chip development\cite{hi5_sanny}.

\subsection{Cryogenic nuller prototype}

\begin{figure}[!h]
\begin{center}
\begin{tabular}{c}
\includegraphics[width=0.97\textwidth]{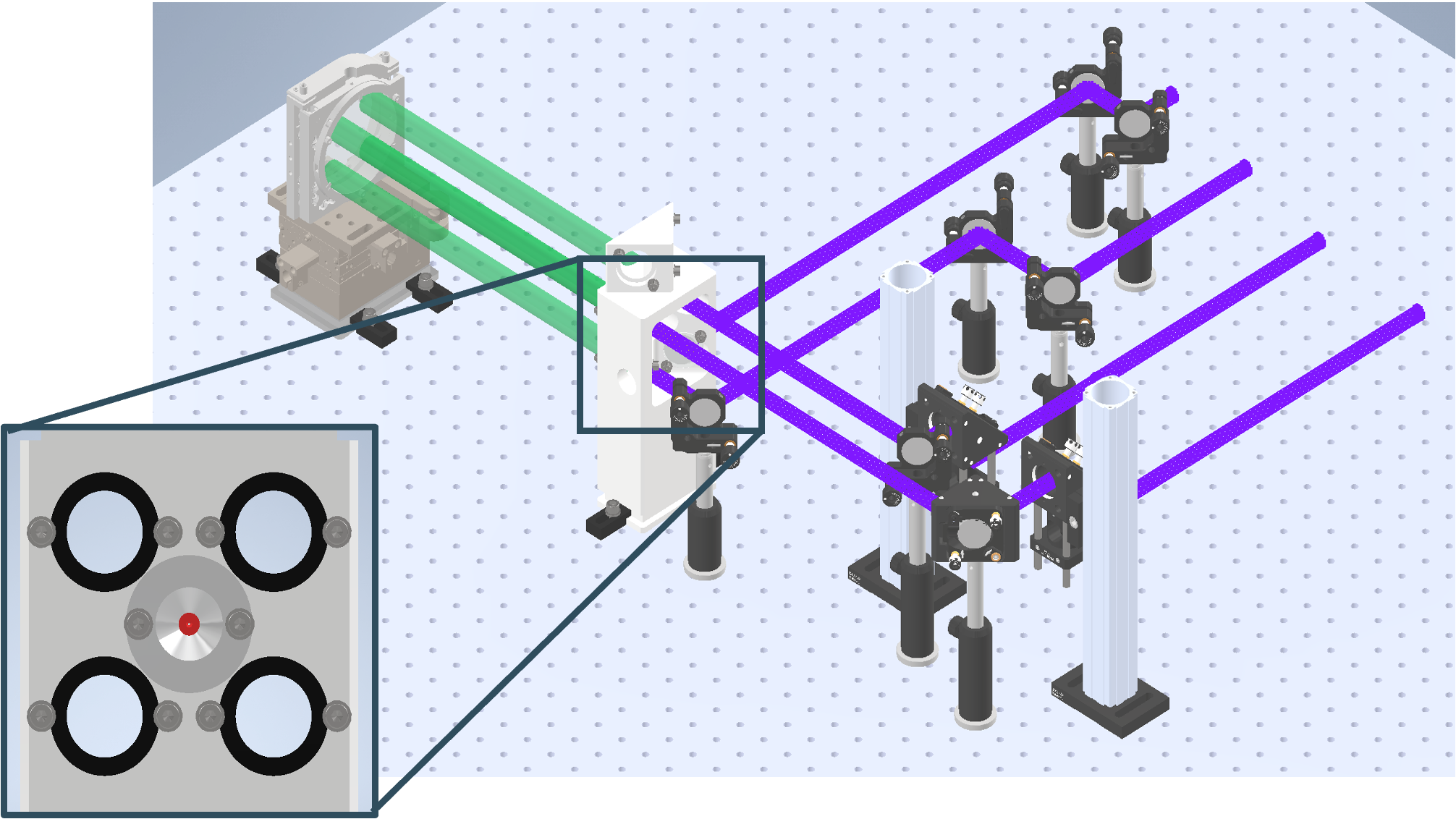}
\end{tabular}
\end{center}
\caption{\label{Fig:vlti_lab} Design of the VLTI beam simulator developed and assembled at KU Leuven. It is a fully reflective design compatible with the wide wavelength range needed for Asgard.}
\end{figure}

\begin{figure}[!h]
\begin{center}
\begin{tabular}{c}
\includegraphics[width=0.57\textwidth]{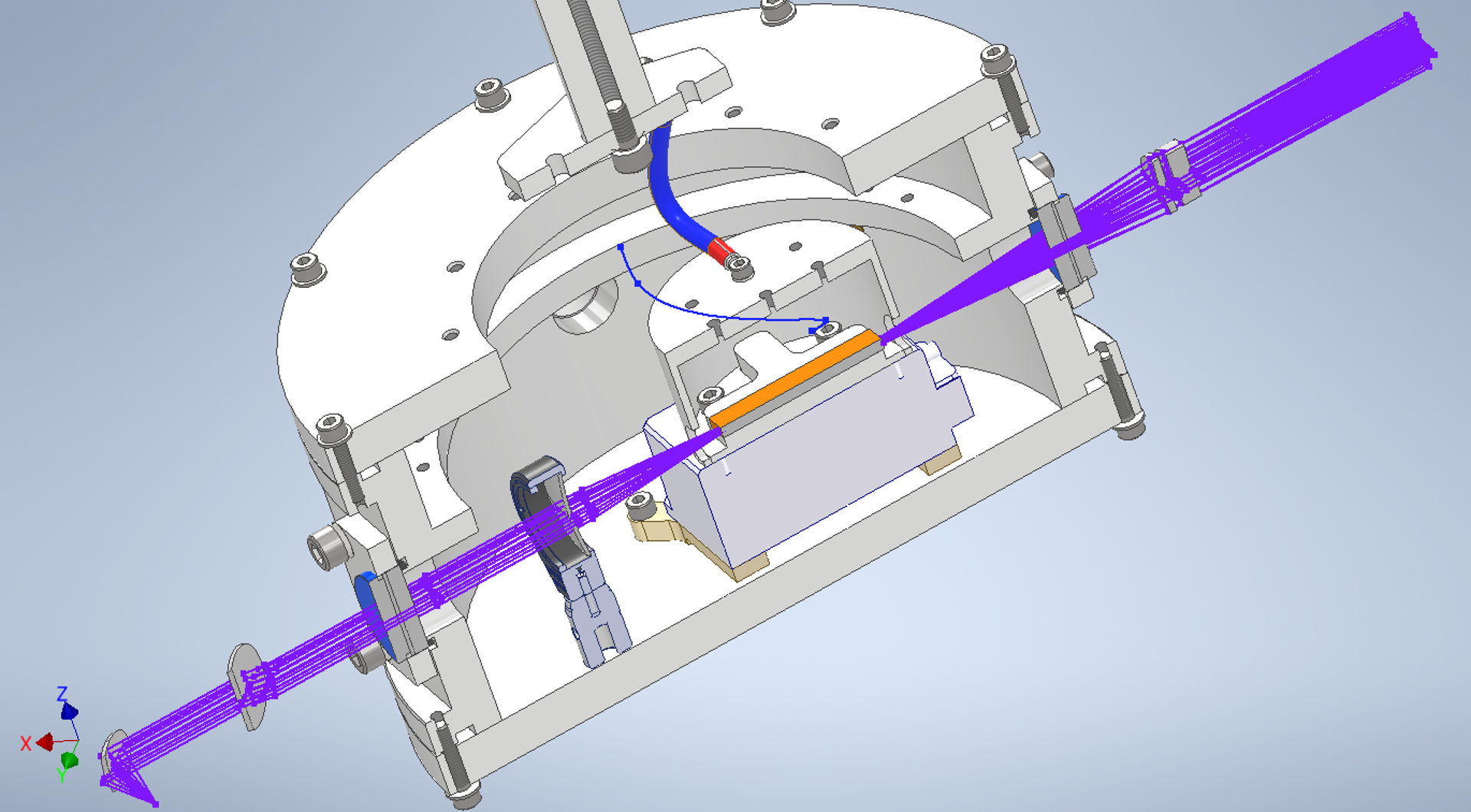}
\includegraphics[width=0.41\textwidth]{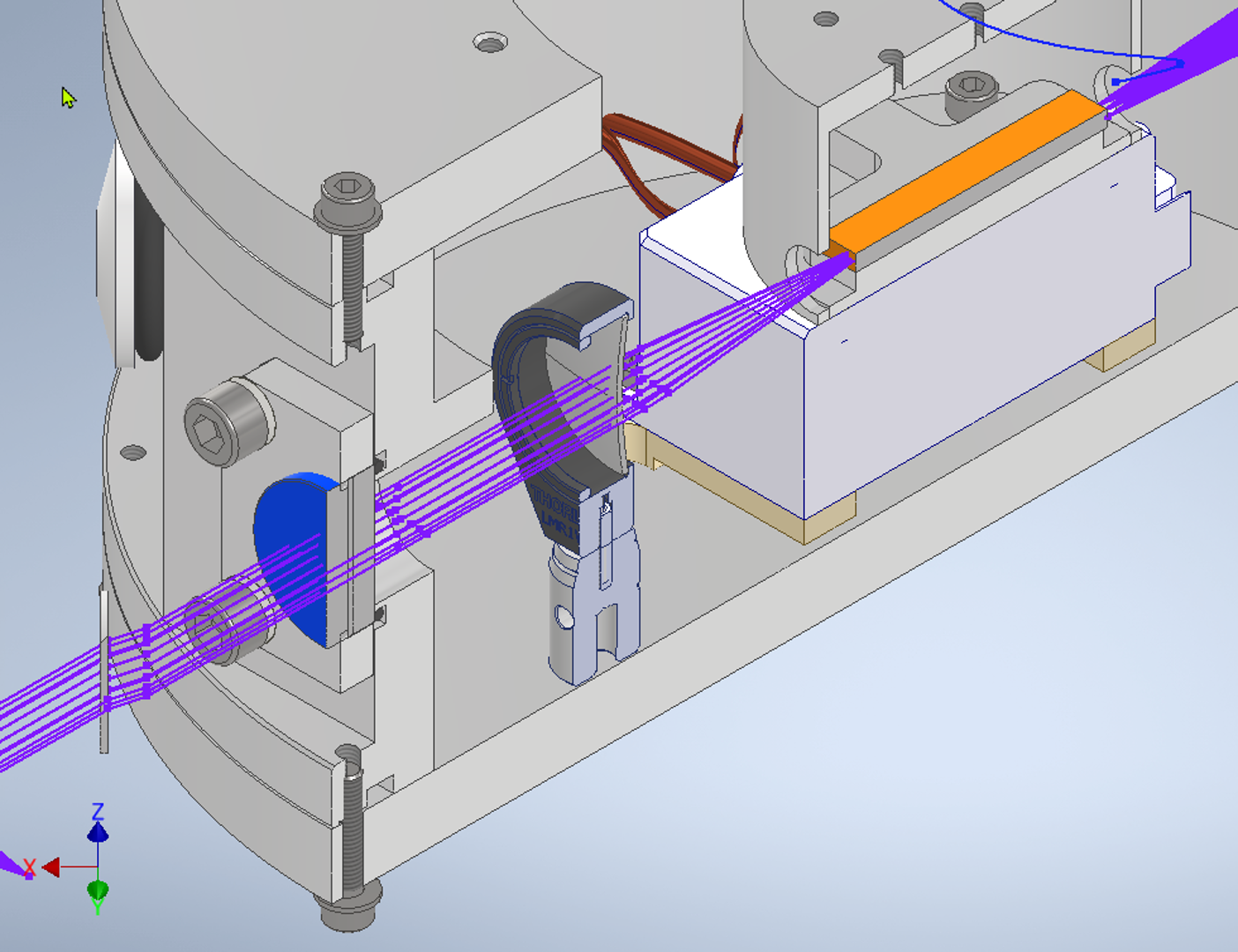}
\end{tabular}
\end{center}
\caption{\label{Fig:cryostat_test} Current design of the test cryostat being assembled in KU Leuven to test cold optics, including the photonic chip, and cryo piezo stages.}
\end{figure}

Figure~\ref{Fig:vlti_lab} shows the design of the VLTI beam simulator developed and assembled in the laboratory at KU Leuven. Four plane wavefront sub-pupils are selected after a parabola-collimated beam from a single mode fiber.  With the exception of the SM fiber, it is a fully reflective design compatible with the wide wavelength range needed for Asgard. Figure~\ref{Fig:cryostat_test} shows the current design of the test cryostat developed for testing components and sub-systems at cryogenic temperatures (photonic chip, lenses, coatings, and cryo motion stages). First tests of the cold stage, photonic chip, and injection procedure are scheduled for September 2022. 

\section{Expected performance}

\begin{figure}[!h]
    \centering
    \includegraphics[width=0.45\textwidth]{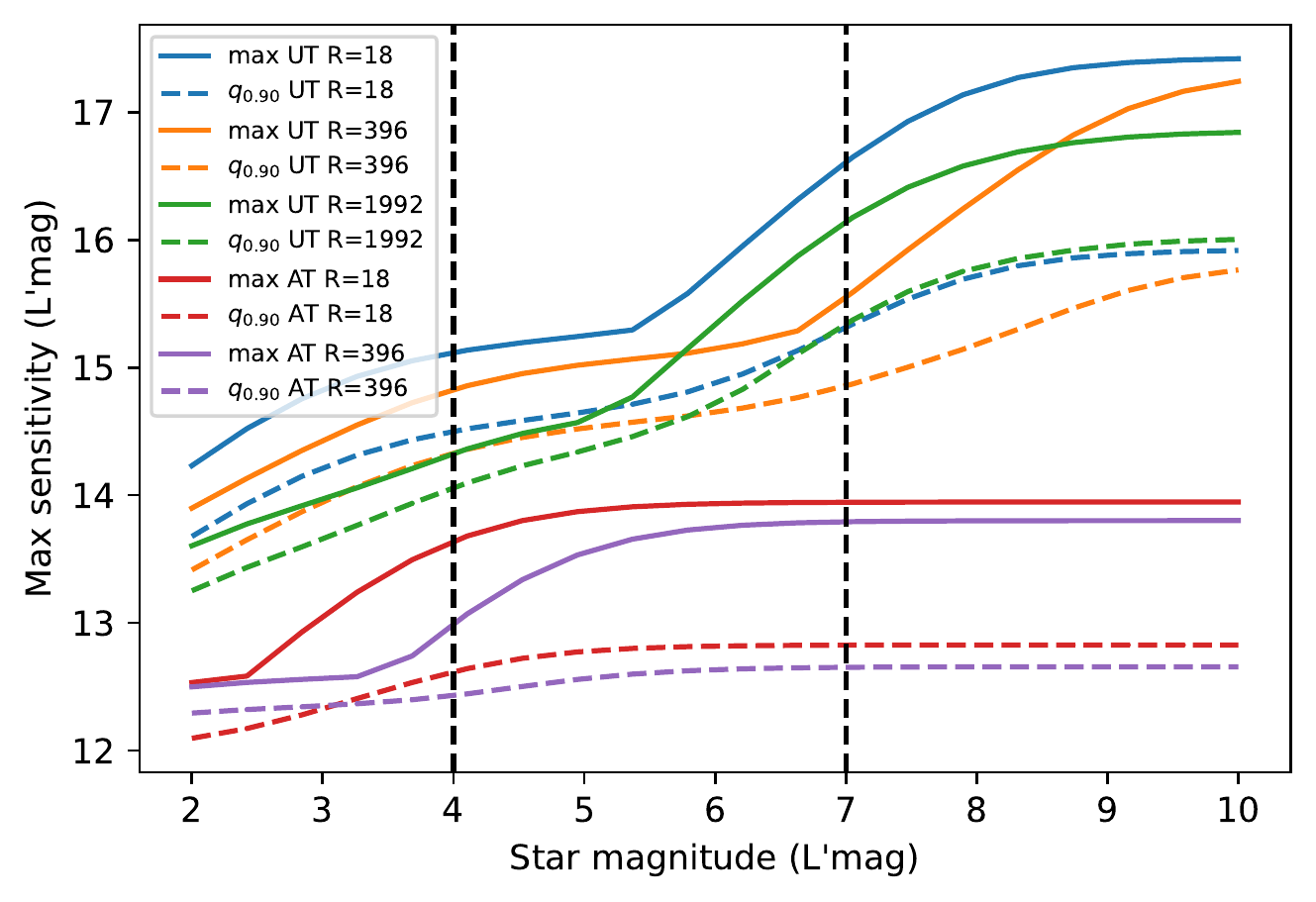}
    \includegraphics[width=0.45\textwidth]{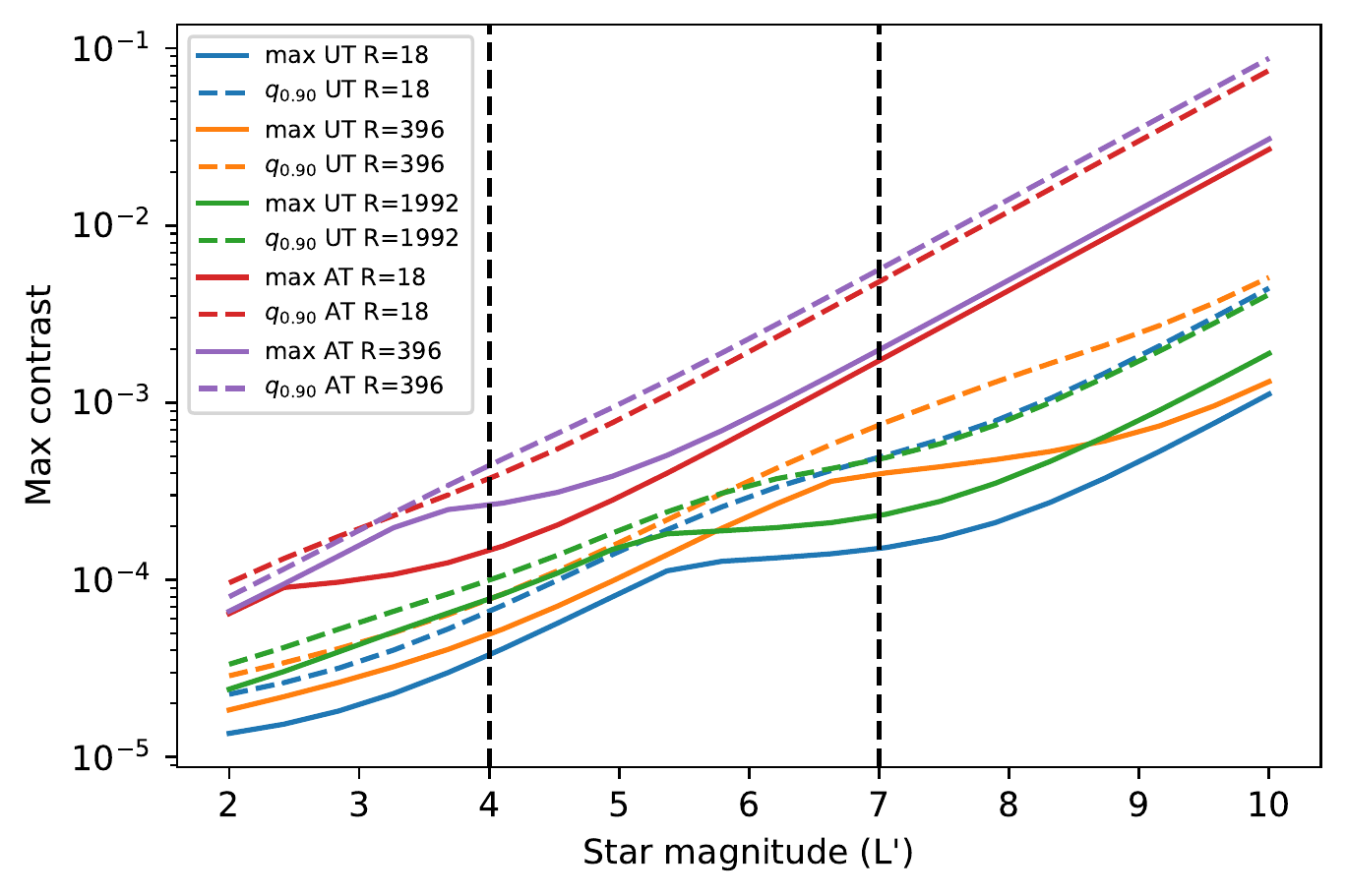}
    \caption{A comparison of the sensitivity (left) and contrast (right) for different instrumental configurations as a function of the target magnitude. Plain lines are the maximum pixel of the map and the dashed line is used to show a 0.90 quantile, which covers a more useful fraction of the field of view. These simulations assumes OPD residuals of 100\,nm RMS for both ATs and UTs.}
    \label{fig:comparison}
\end{figure}

In order to investigate the instrumental performance of Hi-5 and make instrumental trade-offs, we have developed a new end-to-end simulator (SCIFYSim). The code can now be found on GitHub\footnote{https://github.com/rlaugier/SCIFYsim}. Figure ~\ref{fig:comparison} illustrates the performance estimates (sensitivity and contrast) for different VLTI configurations and spectral resolutions. Both the value of the most sensitive pixel an of the $q_{0.9}$ quantile (for given angular separation) are plotted. Such plots must be interpreted with caution, as the sampling and extent of the map may have some influence on the results. On the fainter stars, the smallest separation first peak is a most sensitive region, as indicated by an offset between ($>$3-4 for ATs and 6-7 for UTs), the $q_{0.9}$ is close to the max, indicating a background dominated regime, for which the most sensitive region is the smallest separation first peak. On the brighter stars, the instrumental noise becomes dominant, reducing the sensitivity of this first peak, and the maximum is further away, as highlighted by the dashed line and plain line being closer together. These simulations assume OPD residuals of 100\,nm RMS for both ATs and UTs. While regularly achieved on ATs\cite{lacour_gravity_2019}, achieving such performance on UTs is the subject of ongoing work within the GRAVITY+ collaboration\cite{hi5_bigioli,hi5_Courtney-Barrer}. More information on the performance simulations and future work can be found in a dedicated paper in these proceedings \cite{hi5_laugier}. 

\section{Summary and future work}

Hi-5 is the L’-band (3.5-4.0 $\mu$m) high-contrast imager of Asgard, an instrument suite in preparation for the visitor focus of the VLTI. It was recently submitted to ESO (P110, March 2022) as a first official step to become a VLTI visitor instrument. We present in this paper an update of the science case of Hi-5. We also present the current design of the warm optics, cold optics, and nulling cryogenic test prototype, currently being assembled in the laboratory at KU Leuven. We also present full end-to-end performance based on the latest GRAVITY+ specifications of the VLTI, confirming the potential to reach contrast levels required to detect young exoplanets around nearby bright stars. In the short term, future work will focus on the design of the cryostat for the final instrument (ETH collaboration), test of key components at cryogenic temperatures with the test cryostat, optimization of the nulling chip, and operation/characterization of the HAWAI 2RG\textregistered\,\,\,(procured in Q1 2022).

\acknowledgments 
SCIFY has received funding from the European Research Council (ERC) under the European Union's Horizon 2020 research and innovation program (grant agreement CoG - 866070). This project has received funding from the European Union’s Horizon 2020 research and innovation programme under grant agreement No 101004719. S.K. acknowledges support from an ERC Consolidator Grant (``GAIA-BIFROST'', grant agreement No.\ 101003096). We are grateful for the kind support and constructive interactions with colleagues at ESO, in particular Frédéric Gonte, Xavier Haubois, Antoine Mérand, Nicolas Schuhler, and Julien Woillez.

\bibliography{zotero,report} 
\bibliographystyle{spiebib} 

\end{document}